\documentstyle[prl,aps,preprint]{revtex}

\def\upleftarrow#1{\overleftarrow{#1}}   
\def\uprightarrow#1{\overrightarrow{#1}}  
\def\thru#1{\mathrel{\mathop{#1\!\!\!/}}}
\begin{document}
\tighten

\title{{GAUGE-INVARIANT DECOMPOSITION OF \\ NUCLEON SPIN 
  AND ITS SPIN-OFF}
\thanks {This work is supported in part by funds provided by the
U.S.  Department of Energy (D.O.E.) under cooperative agreement
\#DF-FC02-94ER40818.}}

\author{Xiangdong Ji}

\address{Center for Theoretical Physics \\
Laboratory for Nuclear Science \\
and Department of Physics \\
Massachusetts Institute of Technology \\
Cambridge, Massachusetts 02139 \\
and \\
Institute for Nuclear Theory \\
University of Washington \\
Seattle, Washington 98195 \\
{~}}

\date{MIT-CTP-2517 ~~~ Hep-ph/yymmddd ~~~ March 1996}

\maketitle

\begin{abstract}
I introduce a gauge invariant decomposition of the 
nucleon spin into quark helicity, quark 
orbital, and gluon contributions. The total quark 
(and hence the quark orbital) contribution is 
shown to be measurable
through virtual Compton scattering in a special kinematic 
region where single quark scattering dominates. This 
deeply-virtual Compton scattering (DVCS) has much potential
to unravel the quark and gluon structure 
of the nucleon. 

\end{abstract}
\pacs{xxxxxx}

\narrowtext

The spin structure of the nucleon reflects 
interesting non-perturbative physics in
Quantum Chromodynamics (QCD). 
>From the recent data on polarized deep-inelastic
scattering\cite{e143}, one finds that
about $20 \pm 15$\% of the nucleon spin is 
carried by quark spin or helicity \cite{forte}. 
Natural questions are then where is the remainder of 
the nucleon spin? How to measure or calculate it?
This Letter attempts to provide an answer to them. 

Intuitively, the candidates for the ``missing" spin
are the quark and gluon 
orbital angular momenta
and gluon helicity. In QCD, they can be identified 
with matrix elements of certain quark-gluon 
operators in the nucleon state\cite{jaffemanohar}. 
The problem, however, is that these 
operators take free-field expressions
and are not gauge-invariant in an interacting 
gauge theory. Hence it is doubtful that
their matrix elements have any experimental
significance, although they can be calculated in theory, 
for instance on a lattice, with a fixed gauge.

In this Letter I show that there exists a gauge
invariant decomposition of the QCD angular momentum 
operator into quark and gluon contributions. 
The quark part can be separated further into
the usual quark helicity plus the {\it gauge-invariant 
orbital} contribution. There exists, however, no 
gauge-invariant separation of the gluon part into 
helicity and orbital contributions, although
high-energy scattering favors such 
a separation in the light-like gauge and infinite momentum
frame. The gauge-invariant 
quark and gluon contributions to the nucleon spin 
is shown to asymptotically approach ratio $16:3n_f$, 
where $n_f$ is the number
of active fermion flavors. This result 
is incidentally the same as what Hoodbhoy, 
Tang and this author have derived in a gauge 
non-invariant formulation\cite{jitang}. The
gauge-invariant expression for the angular momentum 
operator allows one to calculate 
meaningfully fractions 
of the nucleon spin carried by quarks and gluons. 
Furthermore, it allows them to be measured 
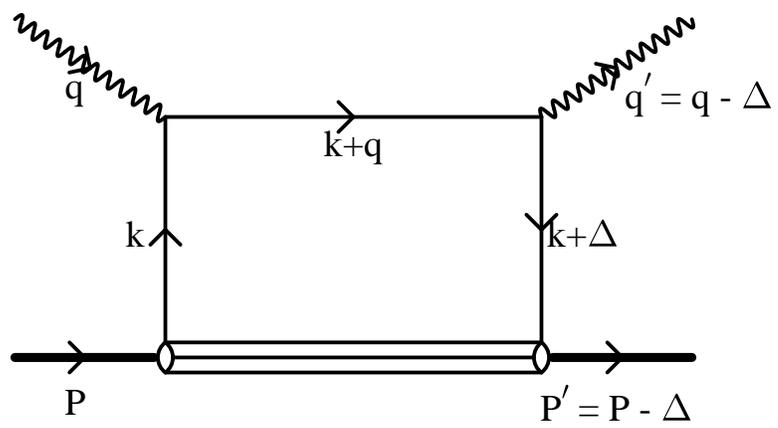
in deeply-virtual Compton scattering (DVCS)
in which the virtual photon momentum approaches the
Bjorken limit. DVCS gives an access to a new class of
nucleon observables---the off-forward parton distributions---which 
are a generalization of ordinary parton distributions
and elastic form factors. Therefore, effectively DVCS provides 
a new ground to explore the quark and gluon structure 
of the nucleon. 

The angular momentum operator in QCD is defined
according to the generators of Lorentz transformation,
\begin{equation}
     J^i = {1\over 2}\epsilon^{ijk} 
      \int d^3x M^{0jk} \ , 
\end{equation}
where $M^{0ij}$ is the angular momentum density, expressible
in terms of the energy-momentum tensor $T^{\mu\nu}$ through
\begin{equation}
      M^{\alpha\mu\nu} = T^{\alpha\nu} x^\mu - 
      T^{\alpha\mu} x^\nu \ . 
\end{equation}
$T^{\mu\nu}$ has the Belinfante-improved form and is
symmetric, gauge-invariant, and conserved\cite{jackiw}. It 
can be separated into gauge-invariant
quark and gluon contributions, 
\begin{equation}
      T^{\mu\nu} = T^{\mu\nu}_q + T^{\mu\nu}_g \ , 
\end{equation}
where the quark part is,
\begin{equation}
     T^{\mu\nu}_q  = {1\over 2}[\bar \psi \gamma^{(\mu} i
      \uprightarrow{ D^{\nu)}} \psi + 
       \bar \psi \gamma^{(\mu} i
           \upleftarrow {D^{\nu)} } \psi] \ , 
\end{equation}
and the gluon part is,
\begin{equation}
     T^{\mu\nu}_g  = {1\over 4}g^{\mu\nu}
       F^2 - F^{\mu\alpha}F^\nu_{~\alpha} \ ,
\end{equation}    
where $(\mu\nu)$ denotes symmetrization with respect to
$\mu,\nu$ indices. I will ignore the issues of 
gauge fixing and trace anomaly \cite{anomaly}, as they do not 
affect the following discussion. 

For the above equations, one sees that $\vec{J}$
can be written as a gauge invariant 
sum, $\vec{J}_{\rm QCD} = \vec{J_q} + \vec{J_g}$, where
\begin{equation}
     J_{q,g}^i = {1\over 2}\epsilon^{ijk} 
      \int d^3x (T^{0k}_{q,g}x^j - T^{0j}_{q,g}x^k) \ . 
\label{def}
\end{equation}
In pure gauge theory, $\vec{J}_g$
by itself is a conserved angular momentum charge, 
generating spin 
quantum numbers for glueballs. It is clear that
$\vec{J}_q$ and $\vec{J}_g$ are interaction-dependent and
thus differ from the corresponding expressions in 
free-field theory. 

To understand the physical content of the gauge invariant
${\vec J}_q$ and $\vec{J}_g$, one can re-express
them using QCD equations of motion and 
superpotentials\cite{jaffemanohar,jackiw}. 
After some algebra, one finds, 
\begin{eqnarray}
      \vec{J}_q &=&  \int d^3x ~ \psi^\dagger
     [\vec{\gamma} \gamma_5  + (\vec{x}\times i\vec{D})] \psi \ , \nonumber  \\
     \vec{J}_g &=& \int d^3x~ (\vec{x}\times (\vec{E}\times \vec{B}))\ , 
\end{eqnarray}
where color indices are implicit. 
One could have guessed the result with going through the formal 
derivation: The quark angular momentum contains the usual 
quark helicity operator plus a gauge-invariant orbital contribution. 
The gluon angular momentum is constructed from the Poynting 
momentum density $\vec{E}\times \vec{B}$. It is a 
bit surprising though that the simple form works for an 
interacting gauge theory. 

At this point, it is instructive to compare 
the above gauge invariant form of the angular 
momentum operator with the free-field
expression\cite{jaffemanohar}. Because the interaction
between quarks and gluons contains no derivative, one can
write $\vec{J}_{\rm QCD} = \vec{J}_q' + \vec{J}_g'$, where
$\vec{J}_{q,g}'$ are interaction-independent (except 
in the definition of $\vec{E}$),
\begin{eqnarray}
      \vec{J}'_q &=& \int d^3 x ~  \psi^\dagger
     \left[\vec{\gamma}\gamma_5  + 
     (\vec{x}\times \vec{i\partial}) \right] \psi \ ,  \nonumber \\
     \vec{J}'_g &=& \int d^3x ~ \left[ (\vec{E}\times \vec{A}) 
      -  E_i (\vec{x}\times \vec{\partial}) A_i \right] \ . 
\end{eqnarray}
Here each term has a straightforward interpretation:
Both the quark and gluon angular momenta are sum of spin and orbital 
contributions. In presence of gauge interactions,  
$\vec{J}'_q$ and $\vec{J}'_g$, as well as the individual terms 
in them except the quark helicity, are gauge-dependent.
For this reason, it is difficult to find experiments to measure the
other contributions to the nucleon spin. 

The gluon angular momentum 
$\vec{J}_g$ does not admit further 
gauge-invariant decomposition as spin and 
orbital contributions, contrary to the
quark case. This is because the
spin space of gluons coincides with the
ordinary space and time, and therefore under spatial
rotation, the orbital and the spin 
representations of the Lorentz group mix. 
If one disregards the issue
of gauge invariance, $\vec{J_g}$ can be viewed as 
a sum of three terms: two terms in $J_g'$ in Eq. (8) and
an interaction-dependent term $-g\int d^3 \psi^\dagger
\vec{x}\times \vec{A}\psi$. The term
$\vec{S}_g = \int d^3x \vec{E}\times
\vec{A}$ has the simple interpretation as the spin
of gluons in the $A^0=0$ gauge\cite{jaffemanohar,balitsky,jaffe}. 
In the finite momentum frame (IFM), 
the gauge condition becomes $A^+=0$ and the rotational 
generators are constructed from the density $M^{+ij}$. 
The nucleon matrix element of $\vec{S}_g$ in the light-like
gauge and IFM is measurable 
in high-energy scattering. In fact, the first moment of 
the polarized gluon distribution $\Delta g(x)$ gives 
\cite{balitsky,jaffe},
\begin{equation}
          \int^1_0 \Delta g(x) dx ~ 2\vec{S} = \langle PS|\hat O|PS\rangle \ . 
\end{equation}
where $\hat O$ is a gauge-invariant operator which reduces 
to $S_g$ in the $A^+=0$ gauge and IFM . 
To maximally utilize this piece of experimental 
information, one can define the gluon 
orbital angular momentum $\vec{L}_g$ as the difference 
$\vec{J}_g-\vec{S}_g$.  
The nucleon matrix element of $\vec{L}_g$ 
in the light-like gauge and infinite momentum frame 
can be deduced from the matrix elements of $\vec{S}$ and 
$\vec{J_g}$ and might offer some insights on the spin 
structure of the nucleon.

According to Eq. (\ref{def}), the $Q^2$ evolution of the
quark and gluon contributions to the nucleon spin 
is the same as that of matrix elements of 
quark and gluon operators. 
The reason is quite simple: forming spatial 
moments of $T^{\mu\nu}_q$ and $T^{\mu\nu}_g$ does 
not change the short-distance 
singularity of the operators. On the other hand, 
in Ref. \cite{jitang}, it was shown that the matrix 
elements of $\vec{J}'_q$ and $\vec{J}'_g$
have the same leading-log evolution as the quark
and gluon contributions to the nucleon momentum. 
If both results above are consistent, 
the interaction-dependent term, $-g\int d^3 \psi^\dagger
\vec{x}\times \vec{A}\psi$, 
shall not affect the leading-log evolution in the
light-like gauge. Indeed, an explicit calculation 
confirms this. If one defines, 
\begin{equation}
       J_{q,g}(Q^2)~ 2{\vec S} = \langle PS|
      \vec{J}_{q,g}(Q^2)|PS \rangle \ , 
\end{equation}
the leading-log $Q^2$ dependence is simply,
\begin{eqnarray}
       J_q(Q^2) = {1\over 2}{3n_f\over 16 + 3n_f} 
   +  \left({\ln Q_0^2/\Lambda^2\over \ln Q^2/\Lambda^2}\right)^{2(16+3n_f)/(33-2n_f)} \left[
          J_q(Q^2_0)-{1\over 2}{3n_f\over 16 + 3n_f}  \right] \ , \nonumber \\ 
       J_g(Q^2) = {1\over 2}{16\over 16 + 3n_f} 
   + \left({\ln Q_0^2/\Lambda^2\over \ln Q^2/\Lambda^2}\right)^{2(16+3n_f)/(33-2n_f)} \left[
          J_g(Q^2_0)-{1\over 2}{16\over 16 + 3n_f}  \right] \ , 
\end{eqnarray}  
where $n_f$ is the number of quark flavors ($J_q+J_g=1/2$).
As $Q^2\rightarrow \infty$, the partition of 
the nucleon spin between quarks 
and gluons approaches the ratio $16:3n_f$, 
the same as the asymptotic partition of the nucleon 
momentum derived by Gross and Wilczek\cite{gross}. 

The gauge-invariant form of the QCD angular momentum 
operator allows one to meaningfully calculate 
and measure the fraction of the
nucleon spin carried by quarks and gluons. 
Recently, Balitsky and I have estimated
$J_{q,g}$ at the scale of 1 GeV using the
QCD sum rule method\cite{jibalitsky}. To see how they can be 
measured in an experiment, I define the 
form factors of the quark and gluon energy-momentum tensors,
\begin{eqnarray}
      \langle P'| T_{q,g}^{\mu\nu} |P\rangle 
       &=& \bar U(P') \Big[A_{q,g}(\Delta^2)  
       \gamma^{(\mu} \bar P^{\nu)} + 
   B_{q,g}(\Delta^2) \bar P^{(\mu} i\sigma^{\nu)\alpha}\Delta_\alpha/2M \nonumber \\
   &&  +  C_{q,g}(\Delta^2)(\Delta^\mu \Delta^\nu - g^{\mu\nu}\Delta^2)/M   
   + \bar C_{q,g}(\Delta^2) g^{\mu\nu}M\Big] U(P)
\end{eqnarray}
where $\bar P^\mu=(P^\mu+{P^\mu}')/2$, $\Delta^\mu = {P^\mu}'-P^\mu$, 
and $U(P)$ is the nucleon spinor. [An analogous
equation for the full tensor was considered by 
Jaffe and Manohar\cite{jaffemanohar}.] The barred form factor 
arises from non-conservation of the tensor currents.  
Taking the forward limit for $\mu=0$ and integrating
over 3-space, one finds that $A_{q,g}(0)$ give 
the momentum fractions of the nucleon carried by 
quarks and gluons ($A_q(0)+A_g(0)= 1$). 
On the other hand, substituting 
the above into the nucleon matrix element of Eq. (\ref{def}),  
one finds,  
\begin{eqnarray}
      J_{q, g} = {1\over 2} \left[A_{q,g}(0) + B_{q,g}(0)\right] \ . 
\end{eqnarray}
Thus, to find the quark and gluon contributions 
to the nucleon spin, one has to 
measure the $B$-form factor, which
is analogous to the Pauli form factor
for the vector current.
 
Since there is no fundamental probe that couples to the quark
and gluon energy-momentum tensors (the graviton does, but only
to the sum), it appears hopeless to measure the form factors.
On the other hand, $T_{q,g}^{\mu\nu}$ does appear in the
operator product expansion (OPE) for the product of vector
currents $TJ_\alpha(\xi) J_\beta(0)$, and thus $B_{q,g}(0)$
is accessible through deep-inelastic
sum rules, like what $A_{q,g}(0$ is measured. 
The complication, however, is that the 
$B_{q,g}(0)$ term does not contribute to the forward 
matrix element and so the usual inclusive deep-inelastic 
process is useless. A way to get around this is 
to measure the off-forward matrix element
of $TJ_\alpha(\xi) J_\beta(0)$ and extrapolating the
form factors to the forward limit. The natural process
to do this is Compton scattering \cite{brodsky1}. To ensure there
is an OPE, one has to have one of the photons far off-shell
and single-quark scattering dominating the process.
To emphasize this special kinematic region, 
I call the process deeply-virtual Compton
scattering (DVCS). In the remainder of this Letter,
I summarize the main features of this process without
going through any detailed proof.

Consider virtual Compton scattering with a
virtual photon of momentum $q^\mu$ absorbed by 
a nucleon of momentum $P^\mu$, and an outgoing real photon of 
momentum $q'^\mu=q^\mu - \Delta^\mu $, and a recoil 
nucleon of momentum $P'^\mu=P^\mu+\Delta^\mu$. 
The deeply-virtual kinematics refer to $q^\mu$ 
in the Bjorken limit, namely, $Q^2=-q^2\rightarrow \infty$,  
$P\cdot q\rightarrow\infty$, and $Q^2/P\cdot q$ finite. 
To ensure a reliable extrapolation to $\Delta^\mu=0$, 
the components of $\Delta^\mu$ shall be as small as possible. 
However, as will be clear below, the best one can do 
is to constrain them on the order of the nucleon mass. 
Given the kinematics above, it is easy to show that 
the only dominant scattering
subprocess involves a single quark absorbing the deeply-virtual 
photon and subsequently radiating a real photon and 
falling back to the nucleon. 
Other subprocesses are down by at least a factor of
$1/Q^2$ and are negligible. I henceforth concentrate
on the dominant subprocess shown in Fig. 1. 

To calculate the scattering amplitude, it is convenient to  
define a special system of coordinates. 
I choose $q^\mu$ and $\bar P^\mu = 
(P+P')^\mu/2$ to be collinear and in the $z$ direction. 
Introduce two light-like vectors, $p^\mu=\Lambda(1,0,0,1)$
and $n^\mu=(1,0,0,-1)/(2\Lambda)$, with $p^2=n^2=0$, $p\cdot n=1$, 
and $\Lambda$ arbitrary. 
I expand other vectors according to $p^\mu$, $n^\mu$ and transverse 
vectors,
\begin{eqnarray}
           \bar P^\mu &=& p^\mu + (\bar M^2/2) n^\mu \nonumber \\
           q^\mu &=& -\xi p^\mu + (Q^2/2\xi) n^\mu   \nonumber \\
          \Delta^\mu &=& -\xi(p^\mu-\bar M^2/2 n^\mu) + \Delta_\perp^\mu 
\end{eqnarray} 
where $ \bar M^2 = M^2 -\Delta^2/4$ and $\xi = Q^2/(2\bar P\cdot q)$. 
The Compton amplitude $T^{\mu\nu} = i\int d^4y e^{-iq\cdot y}
\langle P'|TJ^\nu(y)J^\mu(0)|P\rangle$, where $\nu$ and $\mu$ 
are the polarization indices of the initial and final photons, is, 
\begin{eqnarray}
         T^{\mu\nu}(P,q,\Delta) 
           &&  =  -{1\over 2} (p^\mu n^\nu+p^\nu n^\mu-g^{\mu\nu})
             \int dx \left({1\over x-\xi/2 + i\epsilon} 
            + {1\over x+\xi/2 + i\epsilon}\right) \nonumber 
              \\ && \times \left[ H(x,\Delta^2,\Delta\cdot n)
                     \bar U(P') {\thru n} U(P)
             + E(x, \Delta^2, \Delta\cdot n) 
                   \bar U(P') {i\sigma^{\alpha\beta}n_\alpha
              \Delta_\beta \over 2M}U(P) \right] \nonumber \\ &&
              - {i\over 2}\epsilon^{\mu\nu\alpha\beta}
                p_\alpha n_\beta
                 \int dx \left({1\over x-\xi/2 + i\epsilon} 
            - {1\over x+\xi/2 + i\epsilon}\right)  \nonumber \\ 
            && \times
           \left[ \tilde H(x,\Delta^2,\Delta\cdot n)\bar U(P')
           \thru n \gamma_5 U(P) + \tilde E(x, \Delta^2, \Delta\cdot n)
                  {\Delta\cdot n\over 2M}\bar U(P')\gamma_5 U(P) \right] \ . 
\end{eqnarray}
Thus only 4 of the 12 helicity amplitudes survive the Bjorken limit 
\cite{guichon}. All photon helicity-flipping and longitudinal
photon amplitudes vanish. $H$, $\tilde H$, $E$ and $\tilde E$ 
are new, off-forward, 
twist-two parton distributions defined through the following light-cone
correlation functions, 
\begin{eqnarray}
 \int  {d\lambda \over 2\pi} e^{i\lambda x}
      \langle P'|\bar\psi(-{\lambda n/ 2})\gamma^\mu
            \psi(\lambda n/2)|P \rangle
        &=& H(x,\Delta^2, \Delta\cdot n) \bar U(P')\gamma^\mu U(P) \nonumber \\
         && + E(x,\Delta^2, \Delta\cdot n) \bar U(P'){i\sigma^{\mu\nu}
             \Delta_{\nu}
          \over 2M}U(P) + ...  \nonumber \\
 \int  {d\lambda \over 2\pi} e^{i\lambda x}
      \langle P'|\bar\psi(-{\lambda n/ 2})\gamma^\mu\gamma_5
            \psi(\lambda n/2)|P \rangle
      & =& \tilde H(x,\Delta^2, \Delta\cdot n) 
        \bar U(P')\gamma^\mu \gamma_5 U(P) \nonumber \\
       && + \tilde E(x, \Delta^2, \Delta\cdot n) \bar U(P')
          {\gamma_5\Delta^\mu
         \over 2M}U(P)
        + ...
\end{eqnarray}
where I have negelected the gauge link and the 
dots denote higher-twist distributions.
>From the definition, $H$ and $\tilde H$ are nucleon
helicity-conserving amplitudes and $E$ and $\tilde E$
are helicity-flipping.

The off-forward parton distributions have
the characters of both ordinary parton distributions and
nucleon form factors. In fact in the limit of
$\Delta^\mu \rightarrow 0$, we have 
\begin{equation}
     H(x,0,0) = f_1(x),~~~ \tilde H(x,0,0) = g_1(x)
\end{equation}
where $f_1(x)$ and $g_1(x)$ are quark and quark helicity
distributions. On the other hand, forming 
the first moment of the new distributions, one gets
the following sum rules, 
\begin{eqnarray}
     \int dx H(x,\Delta^2, \Delta\cdot n) &=& F_1(\Delta^2) \ ,  \nonumber \\
     \int dx E(x,\Delta^2, \Delta\cdot n) &=& F_2(\Delta^2) \ ,  \nonumber \\
     \int dx \tilde H(x,\Delta^2, \Delta\cdot n) &=& G_A(\Delta^2) \ , 
           \nonumber \\ 
     \int dx \tilde E(x,\Delta^2, \Delta\cdot n) &=& G_P(\Delta^2) \ . 
\end{eqnarray}
where $F_1$ and $F_2$ are the Dirac and Pauli form factors
and $G_A$ and $G_P$ are the axial-vector and pseudo-scalar form factor. 
The most interesting sum rule relevant to the nucleon spin is,
\begin{equation}
     \int dx x [H(x, \Delta^2, \Delta \cdot n) +
       E(x, \Delta^2, \Delta \cdot n) ]
     = A_q(\Delta^2) + B_q(\Delta^2) 
\end{equation} 
where luckily the $\Delta \cdot n$ dependence, or $C_q(\Delta^2)$
contamination, drops out. Extrapolating the sum rule 
to $\Delta^2=0$, the total quark (and hence quark orbital) 
contribution to the nucleon spin is obtained. By forming still 
higher moments, one gets form factors of various 
high-spin operators. 

It is important to comment on practical aspects of the
experiment. First of all, from the cross section, 
one finds that $E$ and $H$ can be
measured either in unpolarized scattering, or in electron 
single-spin asymmetry through interference with 
the Bethe-Heitler amplitude \cite{guichon}, or in polarized electron 
scattering on a transversely polarized target. A
detailed examination of various 
possibilities, together with some numerical estimates
will be published elsewhere. Second, the DVCS
cross section is down by an order of $\alpha_{\rm em}$
compared with the deep-inelastic cross section, 
but has the same scaling behavior.   
So the cross section is measurable, but statistics
would be a challenging requirement. The ideal 
accelerator for the experiment is ELFE \cite{elfe}. Finally, the extrapolation
of $\Delta^2$ from order $M^2$ to 0 requires dispersive 
study of the form factors of the tensor currents. 
The $B_q(\Delta^2)$ form factor is dominated by 
resonances in the exotic $1^{-+}$ channel. 

The off-forward parton distributions
can be defined for quark helicity-flip (chiral-odd)
correlations, for higher twists, and for gluons. 
DVCS provides
one process to access to these distributions. There
are other processes one can consider to measure
them. For instance, the diffractive $\rho$ or $J/\psi$ 
production studied recently by Brodsky et al.\cite{brodsky}
can be used to measure the off-forward gluon distributions. 
Thus there is now a new territory to explore 
the quark and gluon structure of the nucleon besides 
the traditional inclusive (parton distributions)
and exclusive (form factors) processes.

\acknowledgements
I thank Ian Balitsky for helping to find independent form factors
of the tensor current, D. Beck for calling my attention to
virtual Compton process, and D. Beck, P. Hoodbhoy, 
R. Jaffe, R. McKeown, A. Nathan for discussions and
encouragement.


\begin{references}
\frenchspacing

\bibitem{e143}
J. Ashman et al., Nucl. Phys. B328 (1989) 1; B. Adeva et al., Phys. 
Lett. B302 (1993); P. L. Anthony et al., Phys. Rev. Lett. 71 (1993)
959; K. Abe et. al., Phys. Rev. Lett. 75 (1995) 25. 

\bibitem{forte}
R. D. Ball, S. Forte, and G. Ridolfi, CERN-TH/95-266, 
Edinburgh 95/556, GeF-TH-9/95, hep-ph/9510449, 1995. 

\bibitem{jaffemanohar}
R. L. Jaffe and A. Manohar, Nucl. Phys. B337 (1990) 509.

\bibitem{jitang}
X. Ji, J. Tang, and P. Hoodbhoy, Phys. Rev. Lett. 76 (1996) 740.  

\bibitem{anomaly}
X. Ji, Phys. Rev. D 52 (1995) 271. 

\bibitem{jackiw}
R. Jackiw, in {\it Current Algebra and Anomaly},
World Scientific, Singapore, 1984.

\bibitem{balitsky}
I. Balitsky and V. Braun, Phys. Lett. B267 (1991) 405.

\bibitem{jaffe}
R. L. Jaffe, Phys. Lett. B366 (1996) 359. 

\bibitem{gross}
D. Gross and F. Wilczek, Phys. Rev. D9 (1974) 980. 

\bibitem{jibalitsky}
I. Balitsky and X. Ji, to be published.

\bibitem{brodsky1}
For a summary of probing the nucleon structure with
Compton scattering in other kinematics regions, see
S. Brodsky, talk presented in the workshop ``CEBAF
at Higher Energy", CEBAF, 1994.

\bibitem{guichon}
P. Kroll, M. Sch\"urmann, and P. A. M. Guichon,
Hep-ph/9507298, 1995. 

\bibitem{elfe}
J. Arvieux and E. de Sanctis, The ELFE project, Italian 
Physical Society Conference Proceeding 44 (1993).  

\bibitem{brodsky}
S. J. Brodsky, L. Frankfurt, J. F. Gunion, A. H. Mueller, M. Strikman, 
Phys. Rev. D50 (1994) 3134.  

\nonfrenchspacing
\end{references}
\end{document}